\def\sysname{AltGeoViz\xspace}

\documentclass{vgtc}                          

\graphicspath{{figures/}{pictures/}{images/}{./}} 

\usepackage{times}                     

\usepackage{booktabs}                  
\usepackage{lipsum}                    
\usepackage{xspace}

\usepackage{soul}

\usepackage{mathptmx}                  
\usepackage{colortbl}

\vgtcinsertpkg

\title{\sysname: Facilitating Accessible Geovisualization}

\author{Chu Li\thanks{e-mail: chuchuli@cs.washington.edu}\\ %
        \scriptsize University of Washington %
\and Rock Yuren Pang \\
     \scriptsize University of Washington %
\and Ather Sharif \\
     \scriptsize University of Washington %
\and Arnavi Chheda-Kothary \\
     \scriptsize University of Washington %
\and Jeffrey Heer \\
     \scriptsize University of Washington %
\and Jon E. Froehlich \thanks{e-mail: jonf@cs.washington.edu}\\ %
     \scriptsize University of Washington %
}

\teaser{
  \centering
  \vspace{-12pt}
  \includegraphics[width=\linewidth]{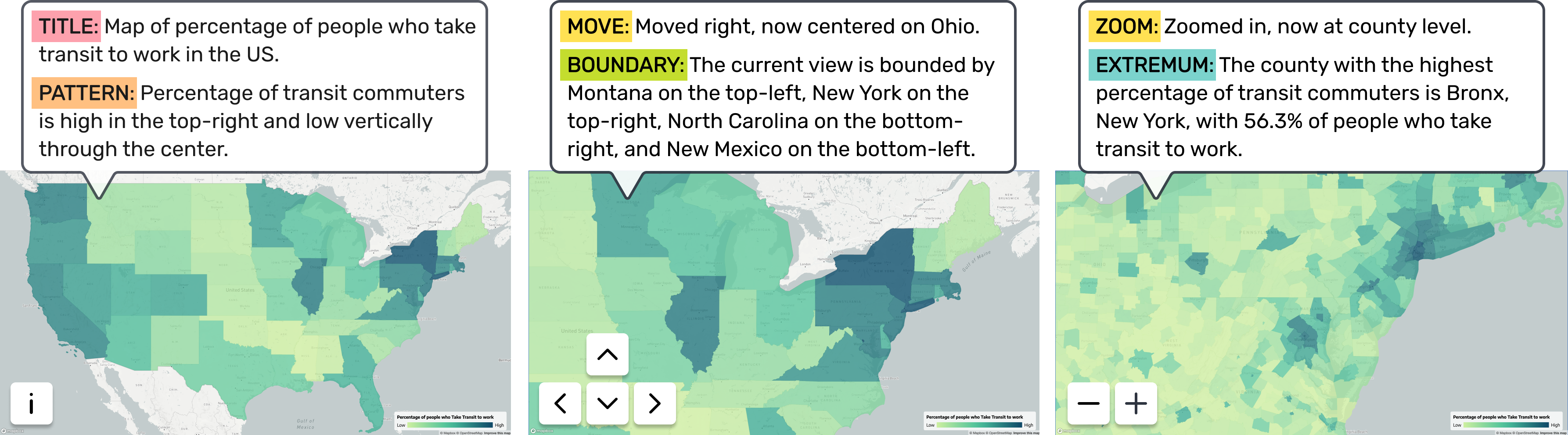}
  \caption{\sysname enables screen-reader users to interact with dynamic geovisualizations. Upon loading, users are presented with the title, a summary of the general spatial pattern, and extrema and average data values (Left). As users move and zoom, the information is updated, and they can hear the boundary of their current viewport (Center). Data can be shown at different geographic units (\textit{e.g.}, state or county level) depending on the zoom level (Right). See video for a demonstration.
}
  \label{fig:teaser}
}

\abstract{
Geovisualizations are powerful tools for exploratory spatial analysis, enabling sighted users to discern patterns, trends, and relationships within geographic data.
However, these visual tools have remained largely inaccessible to screen-reader users.
We introduce \textit{\sysname}, a new interactive geovisualization approach that dynamically generates alt-text descriptions based on the user's current map view, providing voiceover summaries of spatial patterns and descriptive statistics. 
In a remote user study with five screen-reader users, we found that participants were able to interact with spatial data in previously infeasible ways, demonstrated a clear understanding of data summaries and their location context, and could synthesize spatial understandings of their explorations. 
Moreover, we identified key areas for improvement, such as the addition of spatial navigation controls and comparative analysis features.
}


\begin{document}

\definecolor{myQuoteColor}{RGB}{0, 128, 12}
\definecolor{myHighlightColor}{RGB}{0, 128, 12}

\newcommand{\hlc}[2][yellow]{{%
    \colorlet{foo}{#1}%
    \sethlcolor{foo}\hl{#2}}%
}

\newcommand{\myQuote}[1]{{\textit{``#1''}}}
\newcommand{\todo}[1]{\textcolor{red}{#1}}




\maketitle
\begin{table*}[t]
    \centering
    \includegraphics[width=\linewidth]{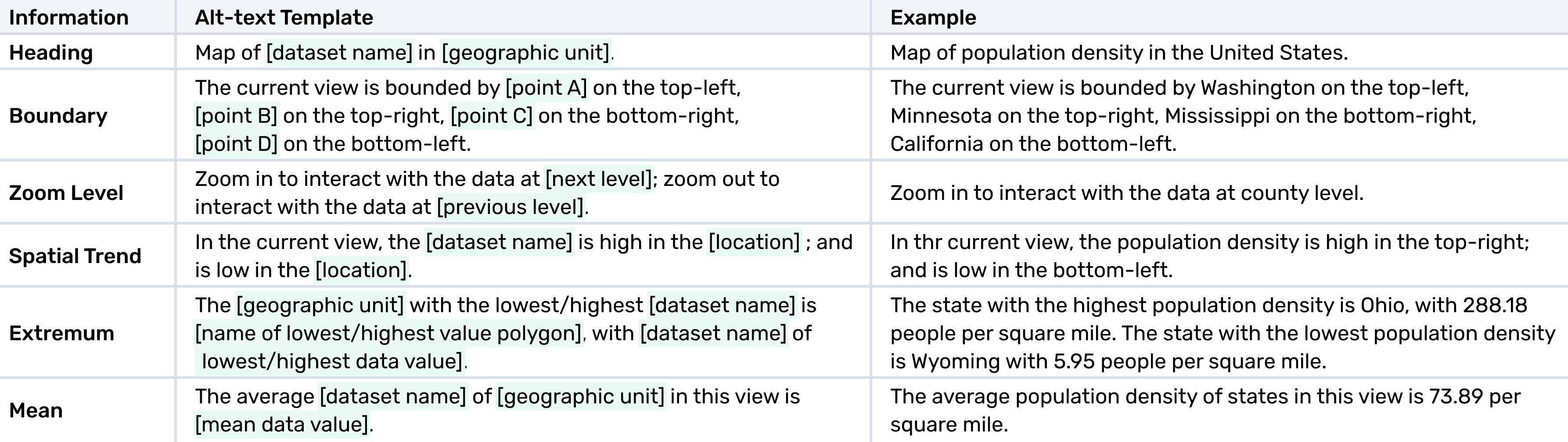}
    \caption{Alt-text template and examples of each component.}
    \label{tab:al-text-content}
\end{table*}

\section{Introduction}
\textit{Geovisualizations} are interactive and dynamic visualizations of geographic information that support exploratory analysis, hypothesis generation, and communicate information~\cite{andrienko_exploratory_2006,coltekin_geovisualization_2018}. 
With improved toolkits~\cite{agafonkin_leaflet_2021, mapbox_mapbox_2024} and broader Internet availability, geovisualizations have become key communicative tools from visualizing COVID occurrences~\cite{times_coronavirus_2020} to election results~\cite{park_extremely_2021}. 
However, geovisualizations are inherently visual, making them inaccessible to screen-reader users unless the visualization creators have explicitly added alt-text~\cite{sharif_understanding_2021,zong_rich_2022}. 
Even then, the alt-text is static, making it impossible for screen-reader users to explore and learn from the data interactively. 
Indeed, a recent study of 15 geovisualizations by Fan \textit{et al.} found significant accessibility issues—none successfully conveyed complex data or higher-level spatial patterns~\cite{fan_accessibility_2023}.

While recent work has introduced new data visualization accessibility methods such as automatic alt-text generation~\cite{mirri_towards_2017, sharif_evographs_2018,sharif_voxlens_2022}, sonification~\cite{ahmetovic_multimodal_2019, apple_inc_audio_2021, sharif_voxlens_2022}, and haptic graphics~\cite{giudice_learning_2012,yu_haptic_2000}, this work is primarily aimed at traditional visualizations (\textit{e.g.,} bar charts, line graphs) rather than geovisualizations. One exception is \textit{Atlas.txt}~\cite{thomas_enabling_2008}, which is a data-to-text natural language generation system that communicates geo-referenced information through screen-readers; however, it does not support interactive explorations.
Other work in static map accessibility introduces techniques to verbally query data points~\cite{sharif_understanding_2022}, provide enhanced navigational structures~\cite{elavsky_data_2023}, and extract data through sonification~\cite{zhao_data_2008}. 

In this paper, we introduce and evaluate new techniques to make \textit{interactive} and \textit{dynamic} geovisualizations accessible to screen-reader users. 
Our custom system \textit{\sysname} auto-generates alt-text based on the user's viewport, enabling screen-reader users to navigate, explore, and extract information from geovisualizations. 
Specifically, \sysname communicates essential information to understand geovisualizations including viewport boundaries, zoom levels, spatial patterns and other summary statistics in any given view. 
To evaluate \sysname, we conducted a remote, qualitative user study with five screen-reader users. Our findings suggest that \sysname enables users to engage with geovisualizations in ways that were previously infeasible: participants effectively identified spatial patterns and other statistical data, and formed spatial understandings of the map based on their explorations.

In summary, we contribute: 
(1) \sysname, an open-source\footnote{GitHub: \href{https://github.com/makeabilitylab/altgeoviz}{https://github.com/makeabilitylab/altgeoviz}} system that auto-generates summaries of interactive and dynamic geovisualizations; (2) Empirical results from a qualitative evaluation of \sysname with five screen-reader users that highlights key strengths and opportunities for improvement.

\section{\sysname Design and Implementation}
Drawing on prior strategies for generating alt-text for geovisualizations ~\cite{thomas_enabling_2008} and other types of visualizations~\cite{jung_communicating_2022,kim_accessible_2021, sharif_understanding_2021}, we describe \sysname's design, including spatial pattern summarization, alt-text generation, and our prototype implementation. Please see our demo video in the supplementary materials.

\vspace{-2pt}
\subsection{Summarizing Spatial Trends}
We summarize spatial patterns by dividing any given map view into a $3\! \times \!3$ grid, then spatially grouping regions with similar values. The $3\! \times \!3$ grid paradigm has previously been employed to assist blind and low-vision users with tasks such as creating graphical information~\cite{kamel_study_2000} and exploring maps through sonification~\cite{zhao_data_2008}.

\begin{figure}[b!]
    \centering
    \includegraphics[width=\linewidth]{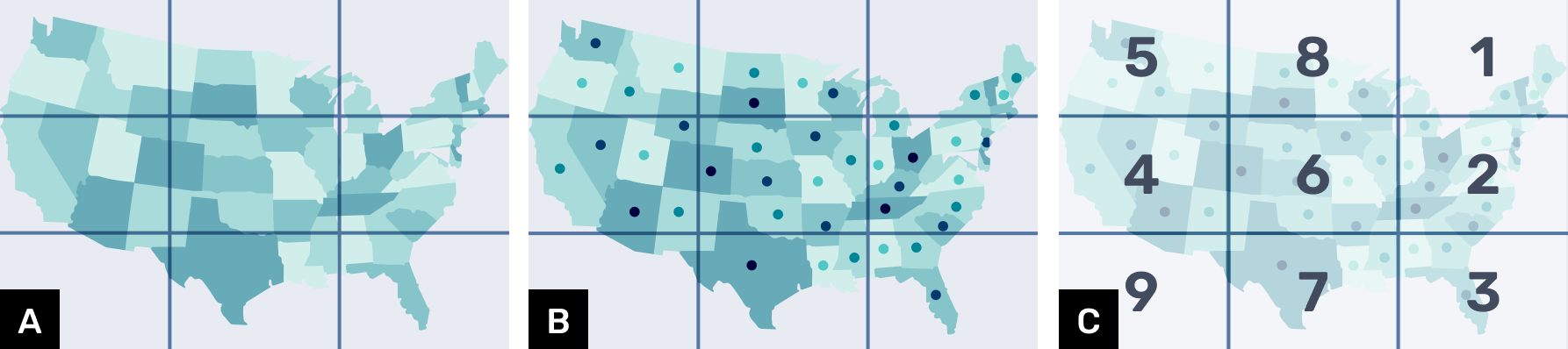}
    \caption{Our grid-based abstraction method for summarizing geovisualization. (A) The bounding box is subdivided into a 3x3 grid. (B) Each polygon is assigned to the grid cell containing its centroid. (C) For each cell, the data values of all associated polygons are aggregated, and the mean value is calculated. Cells are then ranked from 1 (highest) to 9 (lowest).}
    \label{fig:grid-based-abstraction}
\end{figure}

We use a four-step process for grid-based abstraction:
\textbf{(1)} Define a bounding box, which is determined by selecting the smaller of two rectangles: the bounding box of the geographical dataset (in our example, the 48 contiguous U.S. states), or the current viewport of the browser.
\textbf{(2)} Subdivide the bounding box into a $3\! \times \!3$  grid, resulting in nine grid cells, \autoref{fig:grid-based-abstraction}A.
\textbf{(3)} Assign each polygon in the geographic dataset to the grid cell that contains the polygon's centroid, \autoref{fig:grid-based-abstraction}B.
\textbf{(4)} Aggregate within each grid cell the data values across all associated polygons, calculating the mean values, then rank the cells from 1 (highest mean) to 9 (lowest mean), \autoref{fig:grid-based-abstraction}C.
This process allows us to abstract and translate complex visual trends into structured, cell-based summaries.

\begin{figure}[b!]
    \centering
    \includegraphics[width=\linewidth]{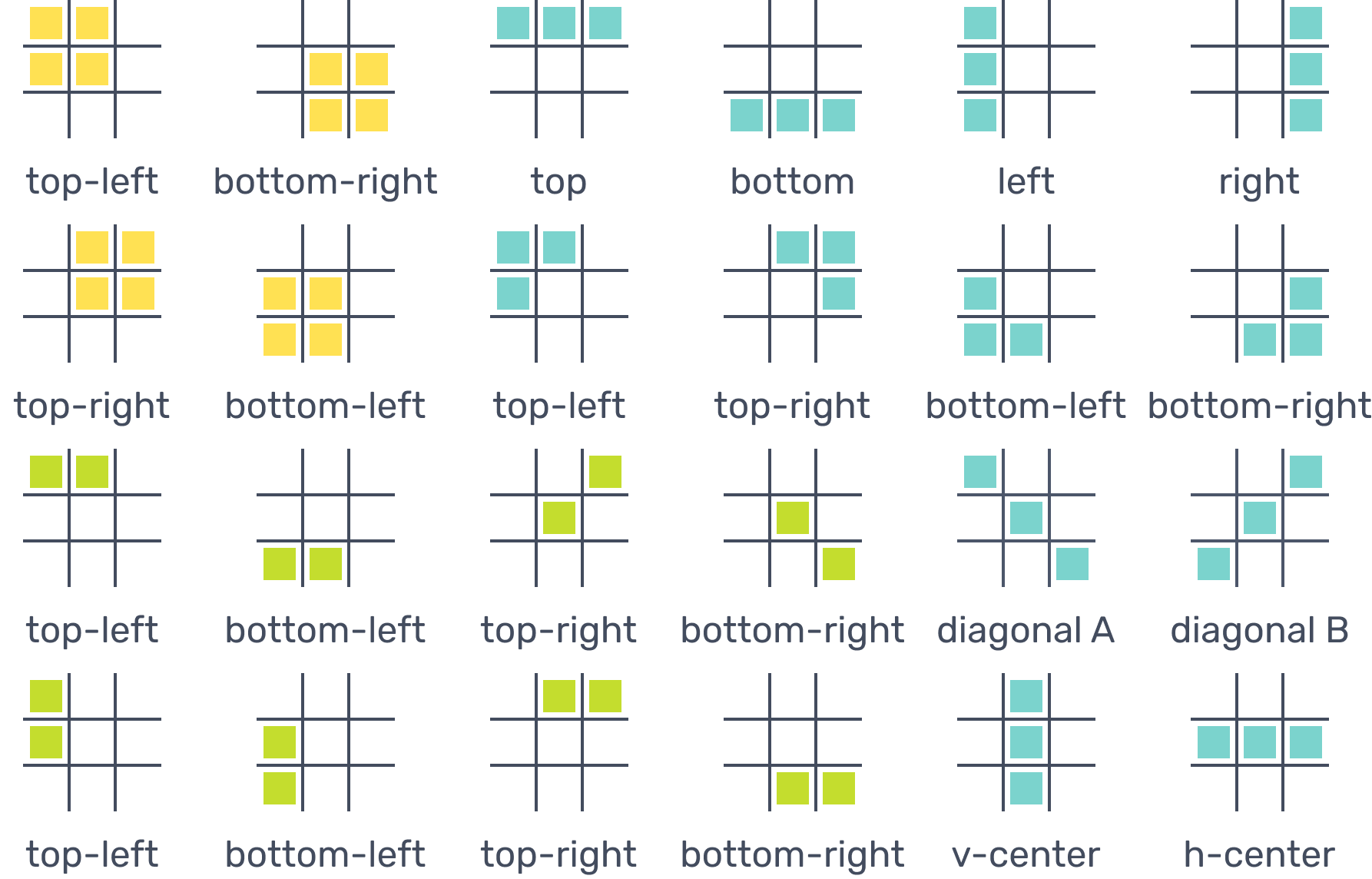}
    \caption{Possible scenarios for adjacent cells in sequential ranking and their corresponding location indicators. Yellow for four, blue for three and green for two sequentially ranked adjacent cells.}
    \label{fig:spatial-grouping}
\end{figure}

To further simplify the grid summary into natural location descriptors, we group grid cells based on value similarity and spatial adjacency. 
This lets the system generate descriptions with terms like ``top-right" instead of cell numbers.
Once the rankings of the grid cells have been established, if four adjacent cells have sequential rankings (\textit{e.g.}, 1, 2, 3, 4 or 6, 7, 8, 9), we classify the data value of this group as either high or low, respectively, and assign the group with a location indicator.
For example, if the four cells in the top-right corner of the $3\! \times \!3$ grid are ranked 1 through 4, we would report the data pattern as high in the top-right corner of the current map view.
If no set of four adjacent cells meet this criterion, then we check groupings of three and subsequently two. 
For these smaller groupings, any combination of three or two numbers within the sequences 1-2-3-4 or 5-6-7-8 would qualify as a grouping.
\autoref{fig:spatial-grouping} lists all possible grouping scenarios and their corresponding location indicators.
If no groupings exist, the system will return: \textit{`no particular regions with high/low [dataset name]'}.

We acknowledge other techniques for summarizing spatial data, such as binning (\textit{e.g. }hexbins)~\cite{setlur_eviza_2016}, kernel density estimation~\cite{okabe_kernel_2009}, and clustering algorithms such as K-means~\cite{okabe_kernel_2009} or DBSCAN~\cite{wang_using_2010}.
Most relevant to our work is \textit{Atlas.txt}~\cite{thomas_whats_2008, thomas_enabling_2008}, which generates descriptions using improved region growing segmentation~\cite{mehnert_improved_1997}.
However, our straightforward techniques provide immediate information retrieval and parameterized outputs that are easily digestible for screen-reader users.
Future work should explore other summarization approaches.


\subsection{Alt-text Design}
Our alt-text design includes the geovisualization description content and interaction instructions.

\textbf{Content.}
Informed by prior work on alt-text content determination for geo-referenced data~\cite{thomas_whats_2008}, and data visualizations more broadly~\cite{jung_communicating_2022,kim_accessible_2021, sharif_understanding_2021}, we design the alt-text template for \sysname to capture essential information of any given viewport, including the title, boundary, zoom level, spatial pattern, extremum, and average values.
\autoref{tab:al-text-content} presents our complete alt-text content design and specific examples for each component.

\textbf{Interaction.}
Prior work emphasizes the importance of initiating alt-text descriptions with an overview of the visualization and providing detailed information only upon request~\cite{kim_accessible_2021,shneiderman_eyes_2003,zong_rich_2022}.
Through iteration and internal testing, we developed an interaction system that balances immediate feedback with cognitive load.
Upon loading \sysname, the screen-reader announces the title of the geovisualization, and users can then interact with the map by pressing the \textbf{m} key. 
For navigation, the \textbf{arrow keys} assist with panning, and the \textbf{+/- keys} with zooming in and out. 
When the navigation keys are pressed, users receive immediate auditory feedback about the action taken and the new center of the viewport, such as \textit{``Moved right, now centered on Missouri."}
For each view, users can press \textbf{i} to hear the spatial pattern and other data values or \textbf{l} to hear the locations of the current viewpoint's four corners.
\autoref{tab:interaction-keys} lists all shortcut keys and their associated behaviors.

\begin{table}[t]
    \centering
    \includegraphics[width=\linewidth]{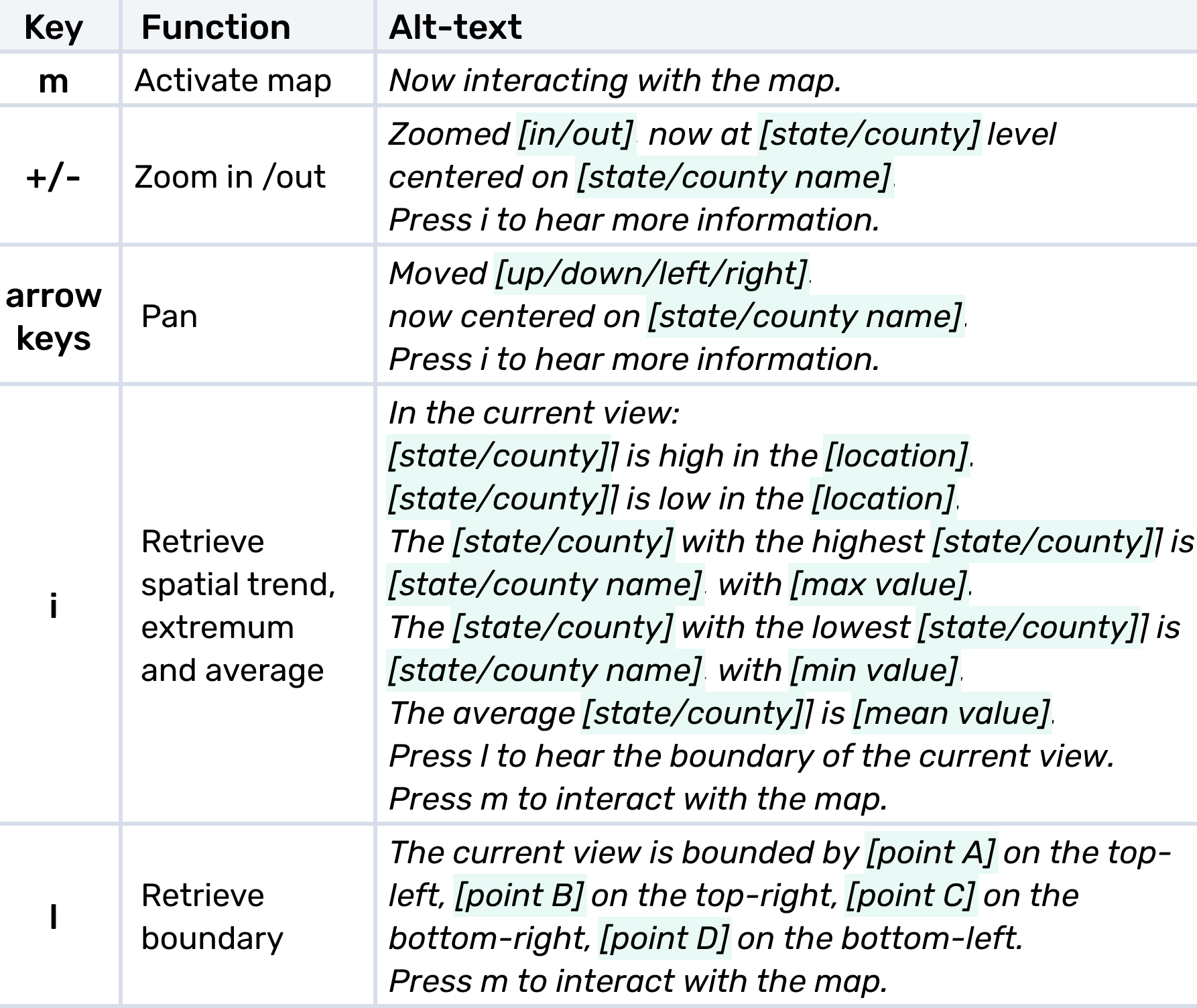}
    \caption{List of all shortcut keys and their associated behaviors.}
    \label{tab:interaction-keys}
    \vspace{-18pt}
\end{table}

\subsection{Implementation}

\sysname's frontend is implemented in MapboxJS, HTML, CSS, and JavaScript, and its backend in Python's Flask framework. We chose DuckDB as the database management system for the geospatial data. \sysname uses Google Cloud for hosting and MongoDB for user log data storage. 
We obtained geographic boundaries for 48 contiguous U.S. states and 3,222 counties from TIGER shapefiles~\cite{bureau_tigerline_2024}. We specifically targeted choropleth (area) maps, future work can extend it to other types, such as symbol maps.
\section{User Evaluation} 

To evaluate \sysname, we conducted a remote user study with five screen-reader users over Zoom.
Participants completed a data exploration task with their chosen screen-reader, after which we asked them questions through a semi-structured interview.
Specifically, we examined if auto-generating and auto-updating alt-text for each map view can support users in exploring and navigating maps effectively (RQ1); 
extracting the spatial patterns and other statistics (RQ2); 
and synthesizing spatial understandings of map explorations (RQ3). 
Additionally, we explored participants' perceptions of overall system usefulness with both Likert scales~\cite{joshi_likert_2015} and interview questions (RQ4).
Participants received a \$30 gift card for one hour of their time.

\subsection{Participants and Procedure}
We recruited five screen-reader users through mailing lists.  
Prior to the study, participants filled out a questionnaire to record their demographic information, screen-reader software, vision-loss level, education level, daily computer usage, and frequency of interacting with online maps (See supplementary materials for participant information table). 
We also provided participants a link to \sysname's tutorial site so the participant could test whether the system's shortcut keys worked with their device.

Study sessions were conducted over Zoom with one to two researchers per session. 
Participants shared their screens when using \sysname and could pause to ask questions at any time.

\textbf{Introduction and Tutorial.} 
Sessions began with a brief overview of the project's motivation and explanation of key concepts related to dynamic geovisualizations. 
To build understanding and comfort with \sysname, participants then used our tutorial website, which featured an interactive choropleth map that displayed population density across the 48 contiguous U.S. states. The map included state- and county-level data drawn from the latest American Community Survey five-year estimates~\cite{bureau_2018-2022_2023}. 
Once participants felt familiar with \sysname, the study task commenced.

\textbf{Study Task.} 
Participants used \sysname to explore a geovisualization similar to the tutorial but with a new dataset of the percentage of the population commuting to work by transit. 
Using \sysname, participants were asked to answer the following: 
(1) Which \textit{U.S. region} has the highest percentage of transit commuters? 
(2) Using pan and zoom, which \textit{county} within that region has the highest percentage of transit commuters,
and what is the percentage value?
(3) Which \textit{state} has the highest percentage of transit commuters, and how does this value compare to the \textit{county} value?

\textbf{Interview.} 
Sessions concluded with semi-structured interviews and 7-point Likert scale questions on usability, perceived usefulness, and mental load (7 was the most positive). We also solicited design ideas for future screen-reader geovisualization systems.

\subsection{Data and Analysis}
All study sessions were audio and video recorded with user consent. Our analysis focused on summarizing high-level themes through qualitative open coding~\cite{charmaz_constructing_2006}. 
One researcher developed a set of themes based on the video transcript and observational notes, then thematically coded the responses.
A second researcher verified these themes~\cite{spall_peer_1998}.
The following participant quotations were slightly modified for concision, grammar, and anonymity.

\section {Findings}
In general, participants responded favorably to \sysname, saying it enabled new interactions with geovisualizations they had not previously experienced. 
Participants could effectively identify spatial patterns and other statistical data, as well as form spatial understandings of maps based on their explorations. 
Nonetheless, participants experienced some challenges in navigating to specific desired locations. We detail key findings and suggestions below.

\subsection{Feedback on \sysname}
We present our findings according to our research questions. Additional details such as task performance and Likert scale results are included in the supplementary material.

\textbf{Exploring and navigating maps effectively (RQ1).}
Four out of five participants (80\%) completed study tasks successfully via panning, zooming, and requesting viewport boundaries and data summaries. Participants remarked on the novelty of the interactive experience, with P2 saying, \myQuote{I have never had such an in-depth interaction with maps before,} and P5 appreciating that \myQuote{the information would change depending on your perspective, which is cool.} We observed that participants enjoyed learning new spatial information and exploring geographic data distributions, particularly when the information aligned with their prior knowledge.

The one participant who did not complete the study tasks had difficulty in navigating to the specific desired location. Participants expressed confusion when panning the map did not change the geographic region at the center of their viewport.
As each keystroke pans a fixed number of pixels, navigating across larger geographic areas often requires multiple keystrokes before centering on a new region.
Most of the participants had not interacted with an interface similar to \sysname before, but their prior experience with data visualizations led them to believe that using arrow keys would directly navigate between discrete geographic regions (P1, P4, P5), rather than positioning a sliding window to continuously pan across the map.
P1 stated that \myQuote{It was frustrating not being able to get to where I wanted to fast.}
P2 and P3 echoed similar sentiments, with P2 mentioning, \myQuote{Getting from area to area [was] a little bit time consuming.} 
Participants also mentioned that when they moved outside areas where data was available, they wanted more guidance on where to navigate next instead of just hearing \myQuote{Currently out of bounds. Please move back on the map.} (P2)

\textbf{Requesting information on location, spatial patterns, and other statistics (RQ2).}
Despite some navigational challenges, participants reported high success in requesting and understanding spatial patterns and data values once they reached the desired location.
All participants felt that it was easy to request data information (\textit{mean=}7; \textit{SD=}0), had a clear understanding of the spatial summaries and other statistics provided (\textit{mean=}6.8; \textit{SD=}0.4), and gave high ratings for how easy it was to understand their location on the map at any given time (\textit{mean=}6.4; \textit{SD=}0.9).
As P3 mentioned, \myQuote{I like that I can find information fairly quickly.} 
P5 appreciated \myQuote{how intuitive the shortcuts were} for interactions and noted that \myQuote{it was really intuitive with which level you were zoomed in on and what your boundaries were.}
P3 echoed this, saying \myQuote{It gave me the boundary information, so it was very easy [to know where I was].}
However, P1 warned their location awareness was straightforward given prior knowledge of U.S. geography but might be challenging for unfamiliar regions. 

\textbf{Synthesizing spatial understandings from explorations (RQ3).}
All participants could describe the spatial paths they traversed through during the tasks, regardless of whether they completed the tasks successfully. 
For example, P1 recalled their navigation process in detail:\myQuote{From the basic view, I needed to go to the upper-right-hand corner quickly and efficiently. So I used my big windows to zoom over, then zoom in. Zoom in to know I am in the upper-right-hand quadrant.} This illustrates how P1 synthesized the directional movements and zoom operations into a cohesive mental representation of navigating to the intended map region.

\textbf{Perceived utility and mental load (RQ4).}
All participants felt \sysname was a positive step towards making geovisualizations usable and accessible for screen-reader users.
P4 spoke to the prevalence of map visualizations online but the lack of screen-reader access: \myQuote{Anytime information needs to be shown, they'll put out a map, even for my children's school. I see them a lot, but what I don't see is a useful summary. Even when there is, it's just a brief overview that doesn't answer my specific questions.}

Given such context, participants provided positive feedback and highlighted how \sysname enabled new interaction experiences with map visualizations.
P2 noted, \myQuote{One of the grad school classes I wanted to take was GIS. But I had to forego taking that class because there was nothing like this [\sysname].}
Participants felt \sysname provided meaningful spatial awareness and perspectives that would have been difficult to achieve through non-interactive formats.
As P1 stated, \myQuote{I really enjoyed being able to navigate the U.S. and just get additional data from a graphical perspective,} and \myQuote{It gives me that kind of spatial awareness...whereas I would not be able to figure out that with a drop-down box.}

When asked about the overall mental demand of using the prototype, participants gave an average rating of 5 out of 7 (\textit{SD}=1.0). Participants explained that this level of mental demand stemmed from the novelty of the interaction itself. As P2 stated, \myQuote{I've never had this in-depth interaction with maps, so it required quite a bit of cognitive load initially.}

\subsection{Suggestions for Improvements}
When asked about potential improvements, participants suggested searchability, comparison capabilities, and complementary information, such as data tables. 

\textbf{Searchability to enhance navigation.} 
To enhance navigation, participants suggested search functions to let them easily and quickly locate specific geographic regions on the map.
P4 stated, \myQuote{I'd like to find a way that I could just get to that specific part of the map quickly.}
P3 agreed stating that \myQuote{I want a searchable area for my map,} and added \myQuote{I just wish I had a way of zooming in on a particular [region].}
P5 proposed that \myQuote{Maybe there can be a little shortcut where you could type in the state abbreviation or the state name, and then [the system] puts you there.}
In addition to search, participants also mentioned other ways to aid navigation.
P1 recommended \myQuote{some instant back-to-home button}, and
P2 proposed the use of \myQuote{one letter navigation.}

\textbf{Additional information.}
Participants requested additional information, including a list of geographical units the view encompassed: \myQuote{I really liked the boundaries, but I wanted more information about what was included in those boundaries or what was outside of those boundaries. Say, there are five states in your view} (P5).
Others suggested an accompanying table available along with the map, \myQuote{I'd like a tabular table with the data in; the map's nice to navigate around, but I want to just select the state of Kansas instead of having to arrow all the way around} (P1).

\textbf{Comparison function.}
The ability to compare statistics between multiple geographic areas was a frequently requested feature. 
P4 stated, \myQuote{It would be nice if I could link [Oregon and Washington] together to get statistics.}
Similarly, P2, inspired by the two datasets in the tutorial and the study task, wanted \myQuote{a comparative analysis between population density and percentage of transit commuters} to explore potential correlations between different datasets.
\section{Discussion and Conclusion}

We presented \sysname, a system that auto-generates and auto-updates summaries of interactive and dynamic geovisualizations for screen-reader users. Our qualitative user study (\textit{N=5}) found that \sysname enabled screen-reader users to interact with geovisualizations in previously infeasible ways.
Participants had clear understandings of spatial patterns, data values and their locations on the map. 
All participants were able to verbalize the pathways they took through exploring the data, which indicates \sysname supported synthesizing spatial understandings.

The study also revealed that participants sometimes struggled with effective navigation using \sysname, likely due to prior mental models shaped by linearly navigating web content~\cite{stockman_influence_2008} rather than 2D map exploration.
Participant feedback highlighted the need for more intuitive navigation capabilities such as text-based search to avoid getting ``lost".
Future iterations should also implement additional user-requested features like comparison across locations/datasets and accompanying underlying data  tables.

A promising future research direction involves developing data-to-text models capable of summarizing key insights from geovisualizations in more sophisticated ways, such as employing different classification models, data aggregation techniques, and clustering algorithms.
Such models have the potential to benefit not just screen-reader users, but also non-screen-reader users. 
Prior research has shown that accurately interpreting geovisualizations like choropleth maps requires some level of geospatial data literacy~\cite{juergens_trustworthy_2020,schiewe_empirical_2019}. 
Factors like color schemes, area sizing, and data classification methods can all introduce perceptual biases that hinder people's ability to accurately discern meaningful spatial patterns from visualizations~\cite{schiewe_empirical_2019}. 
Textual descriptions that communicate underlying data patterns of geovisualizations could enhance screen-reader accessibility while serving as an interpretation aid for a broader audience.

\acknowledgments{
The authors wish to thank Prof. Dan Suciu and Sandy Kaplan. This work was supported by NSF SCC-IRG \#2125087.}


\bibliographystyle{abbrv-doi}

\bibliography{references}


\end{document}